\documentclass[compsoc,conference,a4paper,10pt]{IEEEtran} 
\IEEEoverridecommandlockouts
\usepackage{cite}
\usepackage{amsmath,amssymb,amsfonts}
\usepackage{algorithmic}
\usepackage{graphicx}
\usepackage{textcomp}
\usepackage{bmpsize}
\usepackage{xcolor}
\usepackage{lipsum}
\usepackage{todonotes}
\usepackage[inline]{enumitem}
\usepackage[printonlyused,nolist]{acronym}
\usepackage[colorlinks=true,urlcolor=black]{hyperref}
\def\BibTeX{{\rm B\kern-.05em{\sc i\kern-.025em b}\kern-.08em
    T\kern-.1667em\lower.7ex\hbox{E}\kern-.125emX}}

\usepackage{todonotes}
\usepackage{microtype}
\let\OLDthebibliography\thebibliography
\renewcommand\thebibliography[1]{
  \OLDthebibliography{#1}
  \setlength{\parskip}{0pt}
  \setlength{\itemsep}{0pt plus 0.3ex}
}

\begin{acronym}
	\acro{FL}{Federated Learning}
	\acro{CL}{Collaborative Learning}
	\acro{HFL}{Hierarchical Federated Learning}
	\acro{DP}{Differential Privacy}
	\acro{DOSN}{Decentralized Online Social Network}
	\acro{OSN}{Online Social Network}
\end{acronym}

\begin{document}

\title{Enhancing Privacy via Hierarchical Federated Learning\\
}

\author{\IEEEauthorblockN{Aidmar Wainakh}
\IEEEauthorblockA{\textit{Telecooperation Lab} \\
\textit{Technical University of Darmstadt}\\
wainakh@tk.tu-darmstadt.de}	\\
\IEEEauthorblockN{Tim Grube}
\IEEEauthorblockA{\textit{Telecooperation Lab} \\
\textit{Technical University of Darmstadt}\\
grube@tk.tu-darmstadt.de}	
\and
\IEEEauthorblockN{Alejandro Sanchez Guinea}
\IEEEauthorblockA{\textit{Telecooperation Lab} \\
\textit{Technical University of Darmstadt}\\
sanchez@tk.tu-darmstadt.de}	\\
\IEEEauthorblockN{Max M{\"u}hlh{\"a}user}	
\IEEEauthorblockA{\textit{Telecooperation Lab} \\
\textit{Technical University of Darmstadt}\\
max@tk.tu-darmstadt.de}
}

\IEEEoverridecommandlockouts
\IEEEpubid{\makebox[\columnwidth]{\copyright 2020 IEEE. Personal use of this material is permitted.} \hspace{\columnsep}\makebox[\columnwidth]{ }}

\maketitle

\IEEEpubidadjcol

\begin{abstract}
Federated learning suffers from several privacy-related issues that expose the participants to various threats.
A number of these issues are aggravated by the centralized architecture of federated learning.
In this paper, we discuss applying federated learning on a hierarchical architecture as a potential solution. 
We introduce the opportunities for more flexible decentralized control over the training process and its impact on the participants' privacy.
Furthermore, we investigate possibilities to enhance the efficiency and effectiveness of defense and verification methods. 

\end{abstract}


\section{Introduction}\label{sec:introduction}
In recent years, data breaches have dramatically increased~\cite{McCandless2018}.
At the same time, more services are making use of large amounts of personal data as part of machine learning implementations to provide value to users. 
These two factors have contributed to making users of services increasingly concerned about their privacy.
The concept of \ac{FL} has been proposed to alleviate this issue by allowing multiple users to build a joint model without sharing their data, under the coordination of a central server~\cite{mcmahan2016communication}. 
However, \ac{FL}'s distributed training process, together with its strong dependence on a central server, have increased the attack surface against users yet again~\cite{kairouz2019advances}. 
Fully decentralized learning based on peer-to-peer topology has been proposed to cope with the issues associated with having a central server, such as a performance bottleneck and a single point of failure~\cite{tang2018d}.
However, the convergence time and robustness against user churn remain open challenges.

A hierarchical architecture lies in between the two extremes, centralized and fully decentralized architectures, as a solution able to cope with complexity, scalability, and failure-related issues~\cite{jaques1991praise}.
Applying \ac{FL} based on a hierarchical architecture is referred to as \ac{HFL}.
\ac{HFL} naturally matches emerging decentralized infrastructures (e.g., edge and fog computing) and heterogeneous nature of real-world systems~\cite{lin2018don}.
Most importantly, \ac{HFL} provides a unique capacity for improving privacy. 
As such, it reduces the centralization of power and control in the hands of the central server.
In addition, \ac{HFL} allows flexible placement of defense and verification methods within the hierarchy and enables these methods to be applied more efficiently and effectively.
Furthermore, \ac{HFL} creates the possibility to employ the trust between users to mitigate a number of threats. 

In this paper, we explore the potential benefits of using \ac{HFL} to address privacy-related issues currently discussed in the \ac{FL}-community.  
The rest of this paper is organized as follows. 
In Section~\ref{sec:FL}, we present \ac{FL} and three of its privacy-related issues. 
Then, we introduce, in Section~\ref{sec:HFL}, the concept of \ac{HFL} and potential applications.
In Section~\ref{sec:implications}, we discuss the privacy implications of \ac{HFL}.
Finally, we wrap up with a conclusion section.

\section{Federated Learning and Open Issues} \label{sec:FL}
\acf{FL} is a machine learning setting where multiple entities (users) train a joint model.
Each user trains the model locally on their data and shares only the model updates with a central server.
The central server aggregates the updates from the users into a new global model.
This setting mitigates a number of privacy risks that are typically associated with conventional machine learning, where all training data should be collected, then used to train a model~\cite{kairouz2019advances}.
In spite of the various benefits of \ac{FL}, several privacy-related issues are still open.
A number of these issues are mainly derived from the centralized nature of the \ac{FL} architecture.
Next, we describe three of these issues, namely centralized control, limited verifiability, and constrained defenses. 

\paragraph{\textbf{Centralized control}}
The server in \ac{FL} plays the role of a central coordinator that performs the following core functions:
\begin{enumerate*}[label*=(\arabic*)]
    \item sampling users, i.e., selecting which users participate in the training process,
    \item broadcasting the model and training algorithm, 
    \item aggregating the model updates, and
    \item broadcasting the updated global model.
\end{enumerate*}
Although such a setting limits the computational cost for the users, it places all the control on a single party, the server.
Thus, the server potentially represents a performance bottleneck and a single point of failure.
Furthermore, from a privacy perspective, concentrating such functions on the server leaves the users with limited or no control over the process. 
Consequently, the users have to rely on trusting the server to perform all the functions correctly while maintaining best privacy practices to protect their model updates.
In this respect, if the server were to be malicious or an adversary were to compromise the server, various attacks could be perpetrated against the users, e.g, reconstruction attacks~\cite{zhu2019deep, wang2019beyond}

\paragraph{\textbf{Limited verifiability}}
In \ac{FL}, the server and users perform several computations locally and share the results with each other.
The users share their updates, while the server shares the aggregated model.
In order to allow the server and users to prove to each other that they perform the expected computations correctly and share a legitimate output, the computations need to be verified.  
Multiple approaches based on zero-knowledge proofs have been proposed to apply this verification~\cite{kairouz2019advances}.
However, these approaches suffer from two main limitations.
First, the types of the proofs provided are limited (e.g., range proofs~\cite{bunz2018bulletproofs}).
Second, the time required for verification grows typically exponentially with the number of users~\cite{xu2019verifynet, Burhhalter2019}, which renders these approaches unscalable, and thus unsuitable for large-scale applications.
Another technique to tackle the verification issue is to use a Trusted Execution Environment to perform the computations~\cite{kairouz2019advances}.
However, these environments are not widely available on user devices, especially smartphones.

\paragraph{\textbf{Constrained defenses}}
Although \ac{FL} provides improvements for users' privacy, 
it opens the door for plenty of attacks, which can be applied by both malicious servers and malicious users.
In this work, we focus on privacy attacks, where the attacker aims at inferring information about users.
These attacks occur in two modes: passive and active.

\begin{itemize}[leftmargin=1em]
    \item \emph{Passive attacks}:
    The attacker observes the joint model and periodic updates, which can be used to infer information about other users.
    Shokri et al. \cite{shokri2017membership} introduce a membership inference attack using shadow models, 
    which can be performed in \ac{FL}.
    Melis et al. \cite{melis2019exploiting} propose a property inference attack, leveraging snapshots of the global model.
    Zhou et al.~\cite{zhu2019deep} obtain the training data of a target user from their shared updates (gradients). 
    To mitigate such attacks, several defense techniques can be applied by the server and users.
    The users can perturb their model updates before sharing them by using one or a combination of the following methods:
    \begin{enumerate*}[label*=(\arabic*)]
        \item adding noise (e.g., local differential privacy~\cite{mcmahan2017learning}), 
        \item sharing only a fraction of the updates~\cite{shokri2015privacy}, or 
        \item using the dropout technique in a neural network~\cite{srivastava2014dropout}. 
    \end{enumerate*}
    However, these techniques usually lead to a substantial loss of the model accuracy.
    That is, the trade-off between privacy and accuracy needs to be considered~\cite{kasiviswanathan2011can}.
    On the server side, protecting user privacy requires to break the linkability of the individual users with their updates.
    To achieve this, a secure computation can be performed, e.g, secure aggregation~\cite{bonawitz2017practical} or secure shuffling~\cite{bittau2017prochlo}.
    Besides the efficiency issues these techniques suffer from, they hinder the server from detecting malicious updates, creating by that an attack surface for malicious users.

    \item \emph{Active attacks}:
    The attacker participates in the training process as a user, who maliciously modifies their updates to infer information about other users.
    Hitaj et al.~\cite{hitaj2017deep} propose a reconstruction attack where the attacker provokes the target user to overfit the model on their training data.
    Melis et al.~\cite{melis2019exploiting} present a property inference attack based on multi-task learning. 
    Nasr et al.~\cite{nasr2018comprehensive} use gradient ascent to perform a membership inference attack.
    To detect and mitigate such attacks, several methods can be considered, such as anomaly detectors~\cite{shen2016auror}, median-based aggregation~\cite{yin2018byzantine}, 
    trimmed mean~\cite{yin2018byzantine}, and redundancy-based encoding~\cite{data2019data}.
    However, some of these methods require accessing the individual user updates~\cite{shen2016auror}, which is impeded by secure aggregation.
    In addition, the heavy computations and uncertainty of convergence remain main limitations~\cite{chen2018draco}.
\end{itemize}

\section{Hierarchical Federated Learning} \label{sec:HFL}
\begin{figure}[bp]
\vspace{-15pt}
\centerline{
    \vspace{-7pt}
    \includegraphics[width=0.9\columnwidth]{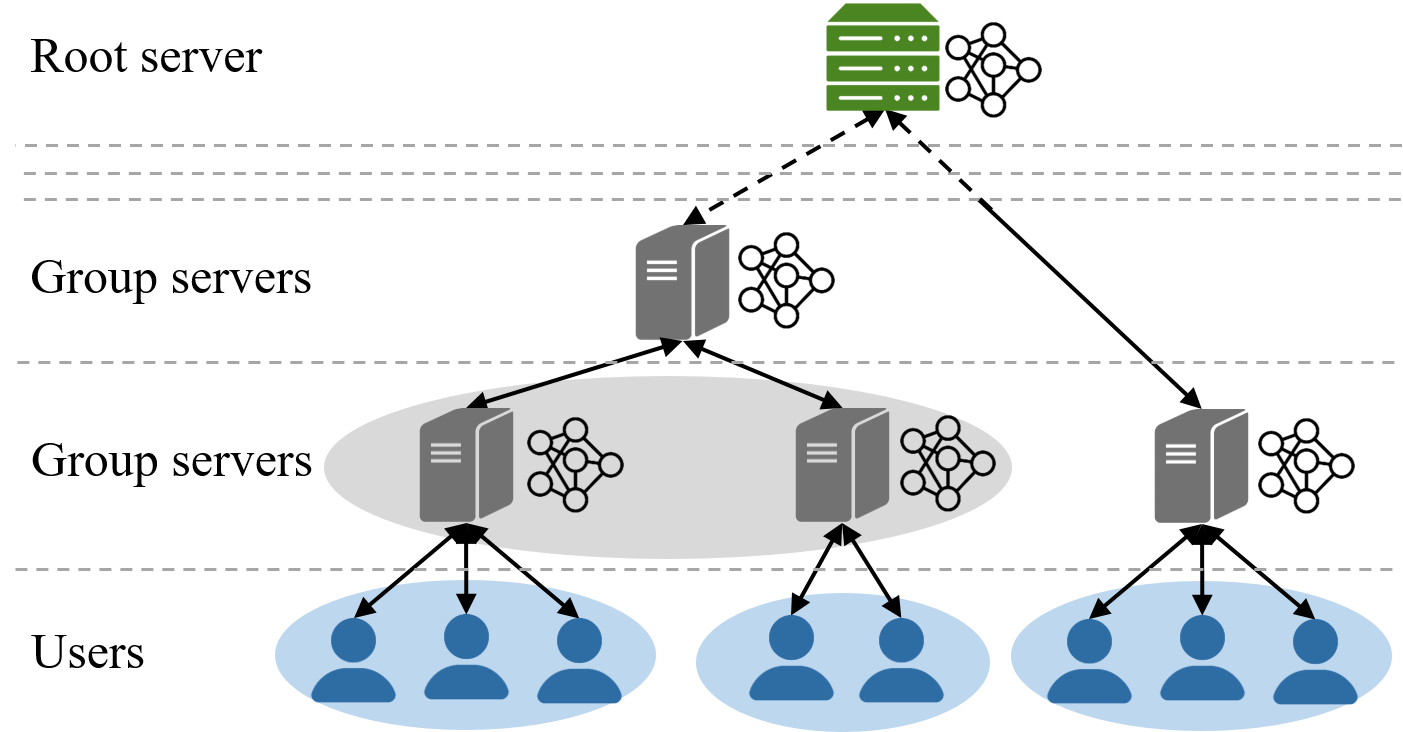}
}
\caption{Example of Hierarchical Federated Learning architecture.}

\label{fig:architectures}
\end{figure}

\ac{FL} is, by definition, coordinated by a central server~\cite{kairouz2019advances}.
We relax this definition to apply \ac{FL} within a hierarchical architecture and refer to it as \emph{\acf{HFL}}.
A restricted version of \ac{HFL}, considering only three layers, has been presented in~\cite{Liu201905, Liu201910, lin2018don}. 
However, privacy was not considered a prime goal.
In \ac{HFL}, there is one \emph{root server}, connected to multiple \emph{group servers}, which are organized in a tree structure (see Figure~\ref{fig:architectures}).
The lowest layer of group servers connects to users (data owners), which are clustered in \emph{user groups}.
The hierarchy can contain multiple layers of group servers and can be unbalanced such that different branches vary in their number of layers.
The users send their model updates to the upper layer to be aggregated by the corresponding group servers.
The aggregation process continues in multiple stages (on each layer) towards the root server.
Every two subsequent layers can have a different number of communication (aggregating) rounds before pushing their models to the upper layer.
After aggregation, the global model is forwarded along the hierarchy downwards to the users.

From an abstract perspective, a hierarchy is a hybrid solution between centralized and fully decentralized architectures.
On the one hand, a hierarchy can cope with the scalability and failure tolerance limitations of centralized architectures.
That is evidenced by the existence and active operation of hierarchical architectures within complex, large-scale systems such as the Internet~\cite{subramanian2002characterizing}.
On the other hand, a hierarchy helps to tackle the management challenges encountered in fully decentralized architectures.
We can observe this advantage in the organizational structure of companies~\cite{jaques1991praise}.


\ac{FL} was originally proposed with the intention to be based on a cloud computing infrastructure~\cite{mcmahan2016communication}.
Such centralized infrastructure represents a network bottleneck, especially with the increase of resource-constrained devices (e.g., smartphones and IoT sensors). 
To alleviate this issue, there is a trend towards decentralized computing infrastructures (e.g., edge and fog computing)~\cite{gedeon2018fog}, which are introducing hierarchies in communication and computing architectures.
\ac{HFL} represents a natural step in extending \ac{FL} into such emerging infrastructures.

Online social networks represent yet another application where the system architecture matches with \ac{HFL}. 
Motivated by privacy-related issues, more decentralized social networks are emerging~\cite{Wainakh2019}.
Diaspora 
and Mastodon 
are two of the most popular, which leverage the concept of local servers.
Users can choose which local server to connect to and where to store their data. 
This architecture represents a hierarchy where some functions are delegated to the higher-layer servers.
In contrast, privacy-related functions (e.g., data storage) are pushed down in the hierarchy to local servers and users.

\section{Privacy Implications of \ac{HFL}} \label{sec:implications}
In this section, we discuss the implications of \ac{HFL} in relation to the privacy-related issues of Section~\ref{sec:FL}.

\paragraph{\textbf{Centralized control}}
In \ac{FL}, the distribution of functions between the server and users is unbalanced;
the control is concentrated on the server.
In contrast, \ac{HFL} allows different \ac{FL} functions to be distributed throughout the hierarchy to the group servers and users, which helps to overcome the performance bottleneck and single point of failure typically found in \ac{FL}.
This is because the group servers can use the local aggregated models within the hierarchy until they converge on the root server at the top. 
Next, we introduce alternative placements for each of the \ac{FL} functions.
\begin{itemize}[leftmargin=1em]
    \item \emph{Sampling users}: 
    In \ac{FL}, the server selects the users to participate in the training process~\cite{mcmahan2016communication}.
    A malicious server could misuse this privilege to attack specific users~\cite{wang2019beyond}.
    In contrast, \ac{HFL} allows sampling to be performed in multiple stages, where a lower layer is sampled by the upper one, till the top.
    This can cope with the gap of trust between the users and server in \ac{FL}, since each two subsequent layers in \ac{HFL} are more likely to have a better base for trust and accountability.
    In fog computing, users can be sampled by edge devices, which can be, for instance, a router in the neighborhood.
    In decentralized social networks, the local server, which is chosen by the users, can perform sampling.
    Also, users can apply privacy-preserving distributed sampling algorithms, such as Anonymous Random Walk used in~\cite{wainakh2019efficient}.

    \item \emph{Broadcasting the training algorithm and model}:
    Congzheng et al.~\cite{song2017machine} have shown that malicious training algorithms can leak information about the training data.
    To address this issue in \ac{HFL}, the group servers can be seen as a protective layer where the algorithms are inspected and verified before being pushed to the users.
    
    From an efficiency perspective, \ac{HFL} provides the possibility for the group servers to adapt the training process to suit the resources of their users' devices, e.g., by adjusting the hyper-parameters of the algorithm or by broadcasting a fraction of the model (sub-model).
    In \ac{FL}, training a sub-model by the users can reveal information about their data to the server~\cite{kairouz2019advances}, while in \ac{HFL} dividing the model to sub-models can happen in different phases through the layers such that an upper layer is not aware how or whether the lower layer divides the model further.

    \item \emph{Aggregating the model updates}:
    As opposed to \ac{FL}, where the aggregation is conducted only by the server, in \ac{HFL}, multiple levels of aggregation occur according to the number of layers in the hierarchy.
    This aggregation scheme implies that only the lowest layer of group servers receive individual updates from users, while all the upper layers process aggregated updates. 
    Consequently, the need for secure aggregation in the upper layers is reduced, which can be interpreted to a performance improvement.
    In addition, the cascade aggregation process helps to protect the identities of the users from the higher-layer servers.

    \item \emph{Broadcasting the updated global model}:
    Some privacy attacks in \ac{FL} require distributing a poisoned model to the target users~\cite{hitaj2017deep}.
    The flat topology at the users level in \ac{FL} allows this type of attacks easy access. 
    While \ac{HFL} provides the possibility to control the model dissemination more effectively through the multiple layers. 
    In case a poisoned model were to be detected within the hierarchy, the group servers could refrain from pushing it to the users.
\end{itemize}

\paragraph{\textbf{Limited verifiability}}
The poor scalability of the verification methods in \ac{FL} (e.g.,~\cite{xu2019verifynet, Burhhalter2019}) hinders their deployment in large-scale applications.
In \ac{HFL}, verifying the updates of the users can be performed by the corresponding group servers.
Grouping users in comparably small groups reduces the computational overhead on the group servers and makes the deployment of verification methods more feasible.

\paragraph{\textbf{Constrained defenses}}
Most of the defense methods in \ac{FL} come with the cost of lower model accuracy~\cite{shokri2015privacy}, considerable computational overhead~\cite{yin2018byzantine}, or both~\cite{data2019data}.
\ac{HFL} provides the possibility to apply these methods in a more flexible manner.
For instance, applying different methods in different parts of the hierarchy, 
such that not all the users suffer from a heavy loss in accuracy nor have to perform costly computations.

As the upper layers in \ac{HFL} process only aggregated models, not individual users' updates, the necessity of secure aggregation can be reduced or even eliminated in these layers.
This can lead to a performance improvement and allows applying anomaly detection methods (e.g.,~\cite{shen2016auror}) to detect malicious models received from lower layers.
Detecting a malicious model from a specific group server can be followed by excluding that server from further aggregation rounds while maintaining the training process functioning normally in the rest of the hierarchy.
Furthermore, the affected server can be notified about its malicious model so it can take more computational expensive countermeasures (e.g.,~\cite{yin2018byzantine}) in their local user group.

Unlike \ac{FL}, \ac{HFL} allows leveraging the trust between users as an additional line of defense.
In social networks, grouping users based on trusted graphs~\cite{jiang2014generating} can reduce the probability of attacks within groups, as it is assumed to be more difficult for malicious users to break into the groups.
Consequently, users may relax their local defenses (e.g., reduce the noise added locally) to achieve a model with higher accuracy.
At the same time, the group servers can take the responsibility of protecting their groups from attacks across groups by applying specific defense methods (e.g., adding noise to the aggregated updates).
\section{Conclusion} \label{sec:conclusion}
In this work, we presented a number of privacy-related issues in \acf{FL}, such as centralized control and constrained defenses.
We discussed how \acf{HFL} can potentially address these issues.
\ac{HFL} utilizes an architecture that lies between the centralized (FL) and fully decentralized learning and comes with a series of privacy advantages.
\ac{HFL} enables high flexibility in functionality distribution, which facilitates known defense and verification methods.
\ac{HFL} allows leveraging the trust between the participants in different application scenarios to mitigate several threats.
Based on this contribution, we wish to encourage further research into the direction of \ac{HFL} to unlock its full potential.



\section*{Acknowledgment}
Funded by the Deutsche Forschungsgemeinschaft (DFG, German Research Foundation) – 251805230/GRK 2050.

\bibliographystyle{plain}
\bibliography{__Bibliography}

\end{document}